\documentclass[12pt]{iopart}
\usepackage{graphicx}
\usepackage{iopams}
\usepackage{dcolumn}
\usepackage{bm}
\usepackage{amsthm}
\usepackage{amssymb}
\usepackage[flushleft]{caption}

\begin{document}

\title{Fidelity susceptibility and quantum Fisher information for
density operators with arbitrary ranks}

\author{Jing Liu$^1$, Heng-Na Xiong$^2$, Fei Song$^1$, Xiaoguang Wang$^{1}$ }

\address{$^1$ Zhejiang Institute of Modern Physics, Department of Physics, Zhejiang University, Hangzhou 310027, China}
\address{$^2$ Department of Applied Physics, Zhejiang University of Technology, Hangzhou 310023, China}
\ead{xgwang@zimp.zju.edu.cn}

\begin{abstract}
Taking into account the density matrices with non-full ranks, we show that the fidelity susceptibility is determined by the support of the density matrix. Combining with the corresponding expression of the quantum Fisher information, we rigorously prove that the fidelity susceptibility is proportional to the quantum Fisher information. As this proof can be naturally extended to the full rank case, this proportional relation is generally established for density matrices with arbitrary ranks. Furthermore, we give an analytical expression of the quantum Fisher information matrix, and show that the quantum Fisher information matrix can also be represented in the density matrix's support.
\end{abstract}


\maketitle

\section{Introduction}

Quantum Fisher information (QFI) is the central concept in quantum
metrology~\cite{c1,c1_1,c1_2,c1_3,c2,c3,c4,c5,c5-1,c6,Joo,Slater}. It
depicts the theoretical bound for the variance of an estimator~\cite{Helstrom,Holevo}
\begin{equation}
\mathrm{Var}(\hat{\theta})\geq\frac{1}{F}.  \label{eq:CRbound}
\end{equation}
Here $\hat{\theta}$ is the estimator for the parameter $\theta$, $\mathrm{Var}(\cdot)$ describes the
variance, and $F$ is
the so-called quantum Fisher information.
A related concept widely used in quantum physics is fidelity, which was first introduced by Uhlmann in 1976~\cite{Uhlmann}. For a parameterized state $\rho(\theta)$ and its neighbor state in parameter space
$\rho(\theta+\delta\theta)$, where $\delta\theta$ is a small change of $\theta$,
the fidelity is defined as
\begin{equation}
f(\theta,\theta+\delta\theta):=\mathrm{Tr}\sqrt{\sqrt{\rho(\theta)}\rho(\theta+\delta\theta)\sqrt{\rho(\theta)}}.
\label{eq:fidelity_definition}
\end{equation}
The form of fidelity is not unique, several alternative forms of fidelity have been proposed and discussed~\cite{fidelity_other,fidelity_other_1,fidelity_other_2}. However, the Uhlmann fidelity is the most well-used form because it has a
natural relation with Bures distance. The fidelity only refers to the Uhlmann fidelity in this paper.
The fidelity in Eq.~(\ref{eq:fidelity_definition}) reveals
the distinguishability between state $\rho(\theta)$ and state $\rho(\theta+\delta\theta)$. It depends on the small change parameter $\delta\theta$.
To avoid this dependence, the concept of fidelity susceptibility (FS) is introduced~\cite{Gu07}. It is  generally believed that the first-order term of $\delta\theta$ in fidelity is zero~\cite{Hubner,Sommers03}, thus FS is determined by the second-order term with the definition
\begin{equation}
\chi_f:=-\frac{\partial^{2}f(\theta,\theta+\delta\theta)}{\partial(\delta\theta)^2}.
\label{eq:fidelity_susceptibility}
\end{equation}
FS is a more effective tool than fidelity itself in quantum physics, especially in detecting the quantum phase transitions~\cite{Gu07,SJGu,SJGu1}.

Interestingly, the above two seemingly irrelevant concepts are in fact closely related to each other. Generally, people vaguely believe that for a given state, the first-order term in fidelity equals to zero and the expression of FS is proportional to that of QFI~\cite{Hubner,Sommers03,Caves94,Twamley}. This is certainly inarguable for the cases with pure states or full-rank density matrices~\cite{Hubner,Sommers03}. However, for density matrices with non-full ranks, a clear and rigorous proof is still lacking. In this work, we will resolve this problem.

Recently, we have obtained the expression of the QFI for a non-full rank density matrix, which is determined by the support of the density matrix~\cite{Liu2013}.
This makes us wonder that if the FS can be written in a similar way and still proportional to the QFI, just like the cases with pure states or full-rank density matrices.
In this paper, we give a detailed calculation of the fidelity for a non-full rank density matrix. We find that its first-order term still equals to zero and FS is also determined by the support of the density matrix.
The whole calculation is rigorous and the expression of FS is proportional to that of QFI.
Our proof can be easily extended to the full-rank case.
In addition, inspired by this result, we further study the quantum Fisher information matrix (QFIM), which is the counterpart of the QFI for the multiple-parameter estimations. Through the calculation, we find that the QFIM is also determined by the support of the density matrix.

The paper is organized as follows. In Sec.~\ref{sec:Fidelity and QFI}, for
a non-full rank density matrix,
we give the detailed calculation of the fidelity. We show that its first-order term also vanishes as the case with full rank. In this way, we get the expression of the FS, which is only determined by the density matrix's support, and proportional to the expression of QFI.
In addition, we apply the expression of QFI (or FS) to a non-full rank X state. In Sec.~\ref{sec:QFIM}, we give the calculation of the QFIM and show that like QFI, QFIM is also determined by the support of the
density matrix. We also apply this expression to a multiple parametrized
X state with non-full rank. Section~\ref{sec:Conclusion} is the conclusion of this work.

\section{Proportional Relationship between FS and QFI\label{sec:Fidelity and QFI}}
In the following, we derive the expression of fidelity for a non-full rank density matrix. From which, we find the first-order term of fidelity vanishes. Then we get the expression of the FS, which is determined by the support of the density matrix. With the corresponding expression of the QFI, we prove the proportional relationship between FS and QFI. Although our proof concentrates on the density matrices with non-full ranks, it could be extended to the ones with full ranks as well.

\subsection{Proof the Proportional Relationship}

We will first obtain the expression of FS from the definition of fidelity in Eq.~(\ref{eq:fidelity_definition}). For brevity, we rewrite the expression of fidelity as  $f=\mathrm{Tr}\sqrt{\mathcal{M}}$ with $\mathcal{M}:=\sqrt{\rho(\theta)}\rho(\theta+\delta\theta)\sqrt{\rho(\theta)}$.
We start our calculation by expanding $\rho(\theta+\delta\theta)$ up to the second order of the small change $\delta\theta$ as
$
\rho(\theta+\delta\theta)=\rho(\theta)+\partial_{\theta}\rho\delta\theta
+\frac{1}{2}\partial_{\theta}^{2}\rho\delta^{2}\theta
$ with $\partial_{\theta}\rho := {\partial\rho}/{\partial\theta}$ and $\partial_{\theta}^{2}\rho := {\partial^{2}\rho}/{\partial\theta^{2}}$.
 Then the matrix $\mathcal{M}$ takes the form
 \begin{eqnarray}
\mathcal{M} = \rho^{2}(\theta)+\mathcal{A}\delta\theta
+\frac{1}{2}\mathcal{B}\delta^{2}\theta.
\end{eqnarray}
where $\mathcal{A}=\sqrt{\rho(\theta)}\partial_{\theta}\rho\sqrt{\rho(\theta)}$ and $\mathcal{B}=\sqrt{\rho(\theta)}\partial_{\theta}^{2}\rho\sqrt{\rho(\theta)}$.
This allows us to assume the square root of $\mathcal{M}$ in the form like
\begin{equation}
\sqrt{\mathcal{M}}=\rho(\theta)+\mathcal{X}\delta\theta+\mathcal{Y}\delta^{2}\theta,\label{eq:XY}
\end{equation}
which is also up to the second-order term of $\delta\theta$.
As a result, taking square of both sides of Eq.~(\ref{eq:XY}), one can find the relations
\begin{equation}
\mathcal{A}  =  \rho \mathcal{X}+\mathcal{X}\rho,\label{eq:X}
\end{equation}
\begin{equation}
\frac{1}{2}\mathcal{B}  =  \rho \mathcal{Y}+\mathcal{Y}\rho+\mathcal{X}^{2}. \label{eq:Y}
\end{equation}
Once the matrices $\mathcal{A}$ and $\mathcal{B}$ are obtained, the information of the matrices $\mathcal{X}$ and $\mathcal{Y}$ will be extracted from these two relationships.

Consequently, the expression of fidelity could be achieved from Eq. (\ref{eq:XY}) as
\begin{equation}
f=1+\mathrm{Tr}(\mathcal{X})\delta\theta+\mathrm{Tr}(\mathcal{Y})\delta^{2}\theta.
\end{equation}
Here $\mathrm{Tr}(\mathcal{X})$ and $\mathrm{Tr}(\mathcal{Y})$ are the first and second order terms of the fidelity, respectively. It is generally believed that the first-order term disappears in fidelity, i.e., $\mathrm{Tr}(\mathcal{X})=0$. However, this conclusion is only well established for pure states or full rank density matrices~\cite{Hubner,Sommers03}. Below we will show that it also holds for density matrices with non-full ranks. This provides a precondition to get the expression of FS, which is determined by
the second-term of fidelity
\begin{equation}
\chi_f=-2\mathrm{Tr}\mathcal{Y}. \label{eq:fidelity_suspt_trY}
\end{equation}

Up to this point, the density matrix $\rho(\theta)$ is still arbitrary, which could be either full rank or non-full rank. Next, to explicitly see the expression of fidelity for non-full rank density matrices, we denote the spectral decomposition of $\rho(\theta)$ as
\begin{equation}
\rho(\theta)=\sum_{i=1}^{M}\lambda_{i}(\theta)|\psi_{i}(\theta)\rangle\langle\psi_{i}(\theta)|.\label{eq:rho_decomposition}
\end{equation}
Here $\lambda_{i}(\theta)$ and $|\psi_{i}(\theta)\rangle$ are the
$i$th eigenvalue and eigenstate of the density matrix, respectively.
$M$ is the rank of the density matrix $\rho(\theta)$, which equals to the dimension of the support of $\rho(\theta)$. We also denote the total
dimension of the density matrix as $N$, which implies that $M\leq N$. In
the following, we will use $\lambda_{i}$, $|\psi_{i}\rangle$ instead
of $\lambda_{i}(\theta)$ and $|\psi_{i}(\theta)\rangle$ for convenience.

It is known that for the density matrix with rank $M=1$ (pure state) or $M=N$ (full-rank), the first order term of fidelity $\mathrm{Tr}\mathcal{X}=0$~\cite{Hubner,Sommers03}. This is a precondition to get the expression of the well-known expression of FS, determined by the second order of fidelity~\cite{Hubner,Sommers03}. However, if one straightforwardly substitutes Eq.~(\ref{eq:rho_decomposition})
into Eq.~(\ref{eq:X}) to get the value of $\mathrm{Tr}(\mathcal{X})$,
one can find the fact that $\langle\psi_{i}|\mathcal{X}|\psi_{j}\rangle$ is arbitrary for $i>M$
and $j>M$, which will result in the arbitrariness of the value of $\mathrm{Tr}(\mathcal{X})$.
That is, $\mathrm{Tr}(\mathcal{X})$ may become undeterminable for density matrices with non-full ranks. This may bring a different expression of fidelity susceptibility.
In fact, this is not true. Below we will show how to avoid this nondeterminacy.

First, we discuss the structure of $\mathcal{A}$ and $\mathcal{B}$. Substituting Eq.~(\ref{eq:rho_decomposition}) into $\mathcal{A}$ and $\mathcal{B}$, and denote $\langle\psi_{i}|\mathcal{O}|\psi_{j}\rangle=\mathcal{O}_{ij}$,
one find that
\begin{eqnarray}
\mathcal{A}_{ij}  &=&  \big{[}\sqrt{\rho(\theta)}{\partial_\theta}\rho\sqrt{\rho(\theta)}\big{]}_{ij}
=\sqrt{\lambda_{i}\lambda_{j}}\left(\partial_{\theta}\rho\right)_{ij}, \nonumber\\
\mathcal{B}_{ij}  &=& [\sqrt{\rho(\theta)}{\partial_\theta^{2}}\rho\sqrt{\rho(\theta)}]_{ij} =\sqrt{\lambda_{i}\lambda_{j}}\left(\partial_{\theta}^{2}\rho\right)_{ij}.
\end{eqnarray}
Here the first and second derivatives of the density matrix are
\begin{eqnarray}
({\partial_\theta}\rho)_{ij}=\lambda_{i}\partial_{\theta}\lambda_{i}\delta_{ij}+\sqrt{\lambda_{i}\lambda_{j}}\left(\lambda_{i}
-\lambda_{j}\right)\langle\partial_{\theta}\psi_{i}|\psi_{j}\rangle,
\end{eqnarray}
\begin{eqnarray}
\left(\partial_{\theta}^{2}\rho\right)_{ij}
& = & \partial_{\theta}^{2}\lambda_{i}\delta_{ij}+2\left(\partial_{\theta}\lambda_{i}
-\partial_{\theta}\lambda_{j}\right)\langle\partial_{\theta}\psi_{i}|\psi_{j}\rangle  \nonumber \\
&  & +\lambda_{j}\langle\psi_{i}|\partial_{\theta}^{2}\psi_{j}\rangle+\lambda_{i}\langle\partial_{\theta}^{2}\psi_{i}|\psi_{j}\rangle
+\sum_{k}2\lambda_{k}\langle\psi_{i}|\partial_{\theta}\psi_{k}\rangle\langle\partial_{\theta}\psi_{k}|\psi_{j}\rangle,
\label{eq:2nd_derivative_rho}
\end{eqnarray}
with $\delta_{ij}$ the Kronecker delta function. From these expressions, one find that when $i>M$ or $j>M$, $({\partial_\theta}\rho)_{ij}=({\partial_\theta}^{2}\rho)_{ij}=0$. That is, both the matrices $[{\partial_\theta}\rho]$ and
$[{\partial_\theta^{2}}\rho]$ are block diagonal with the support dimension of $M$, as well as the density matrix $\rho$. Thus the matrices $\mathcal{A}$ and $\mathcal{B}$
are also block-diagonal ones, with the elements
within the support ($i\leq{M}$ and $j\leq{M}$) are nonzero. As a result, denoting the $M$-dimensional non-zero block of $\rho^{2}$, $\mathcal{A}$
and $\mathcal{B}$
as $\rho_{s}^{2}$, $\mathcal{A}_{s}$ and $\mathcal{B}_{s}$,
we have
\begin{equation}
\mathcal{M}=\left(\begin{array}{cc}
\rho_{s}^{2}+\mathcal{A}_{s}\delta\theta+\frac{1}{2}\mathcal{B}_{s}\delta^{2}\theta & \mathrm{0}_{(N-M)\times M}\\
\mathrm{0}_{M\times(N-M)} & \mathrm{0}_{(N-M)\times(N-M)}
\end{array}\right).
\end{equation}
Since the square root operation on a block diagonal matrix can be manipulated
on each block separately, the square root of $\mathcal{M}$ becomes
\begin{equation}
\sqrt{\mathcal{M}}=\left(\begin{array}{cc}
\sqrt{\rho_{s}^{2}+\mathcal{A}_{s}\delta\theta+\frac{1}{2}\mathcal{B}_{s}\delta^{2}\theta} & \mathrm{0}_{(N-M)\times M}\\
\mathrm{0}_{M\times(N-M)} & \mathrm{0}_{(N-M)\times(N-M)}
\end{array}\right).
\end{equation}

Comparing the above equation with Eq.~(\ref{eq:XY}), one can find that
the matrix $\mathcal{X}$ and $\mathcal{Y}$ must be block-diagonal matrices, of which
only the elements within the support are nonzero.
Therefore, according to Eqs.~(\ref{eq:XY}) and (\ref{eq:X}),
one gets the matrix $\mathcal{X}$ as
\begin{equation}
\mathcal{X}_{ij}=
\left\{
\begin{array}{ll}
\frac{1}{2}\partial_{\theta}\lambda_{i}\delta_{ij}+\frac{\sqrt{\lambda_{i}\lambda_{j}}
\left(\lambda_{i}-\lambda_{j}\right)}{\lambda_{i}
+\lambda_{j}}\langle\partial_{\theta}\psi_{i}|\psi_{j}\rangle, & \hbox{$i,j\in[1,M]$}; \\
0, & \hbox{others}.
\end{array}
\label{eq:X_matrix}
\right.
\end{equation}
Then it is easily found that
\begin{equation}
\mathrm{Tr}(\mathcal{X})=\frac{1}{2}\sum_{i=1}^{M}\partial_{\theta}\lambda_{i}
=\frac{1}{2}\partial_{\theta}\mathrm{Tr}\rho=0.
\label{eq:trX=0}
\end{equation}
Namely, the first order expansion of fidelity vanishes. In this way, the problem of the nondeterminacy of $\mathrm{Tr}({\mathcal{X}})$ is settled. This guarantees that the definition of FS is determined by the second order of fidelity, as shown in Eq.~(\ref{eq:fidelity_suspt_trY}).

Next, to obtain the second order of fidelity, one should know the explicit form
of the diagonal elements of $\mathcal{Y}$. From Eq.~(\ref{eq:2nd_derivative_rho}),
it is easy to get the diagonal elements of $\mathcal{B}$
\begin{eqnarray}
\mathcal{B}_{ii}  =
\lambda_{i}\partial_{\theta}^{2}\lambda_{i}-2\lambda_{i}^{2}\langle\partial_{\theta}\psi_{i}
|\partial_{\theta}\psi_{i}\rangle+\sum_{k}2\lambda_{i}\lambda_{k}|\langle\psi_{i}|\partial_{\theta}\psi_{k}\rangle|^{2}.
\qquad \label{eq:sec_ele}
\end{eqnarray}
where the identity $\langle\partial_{\theta}^{2}\psi_{i}|\psi_{i}\rangle+\langle\psi_{i}|\partial_{\theta}^{2}\psi_{i}\rangle
=-2\langle\partial_{\theta}\psi_{i}|\partial_{\theta}\psi_{i}\rangle$
has been used.
Then according to the relation~(\ref{eq:Y}) and the expressions (\ref{eq:X_matrix}) and (\ref{eq:sec_ele}), one can obtain the diagonal element of $\mathcal{Y}$ within the support as
\begin{eqnarray}
\mathcal{Y}_{ii} = \frac{1}{4}\partial_{\theta}^{2}\lambda_{i}-\frac{1}{8\lambda_{i}}\left(\partial_{\theta}\lambda_{i}\right)^{2} -\frac{1}{2}\lambda_{i}\langle\partial_{\theta}\psi_{i}|\partial_{\theta}\psi_{i}\rangle
+\sum_{k=1}^{M}\frac{2\lambda_{i}\lambda_{k}^{2}}{(\lambda_{i}+\lambda_{k})^{2}}
|\langle\psi_{i}|\partial_{\theta}\psi_{k}\rangle|^{2}. \nonumber\\
\end{eqnarray}

Considering the fact that  $\sum_{i=1}^{M}\frac{1}{4}\partial_{\theta}^{2}\lambda_{i}=\frac{1}{4}\partial_{\theta}^{2}\mathrm{Tr}\rho=0$
and
\begin{equation}
\sum_{i,k=1}^{M}\frac{2\lambda_{i}\lambda_{k}^{2}}{(\lambda_{i}+\lambda_{k})^{2}}|\langle\psi_{i}|\partial_{\theta}\psi_{k}\rangle|^{2}=\sum_{i,k=1}^{M}\frac{\lambda_{i}\lambda_{k}}{\lambda_{i}+\lambda_{k}}|\langle\psi_{i}|\partial_{\theta}\psi_{k}\rangle|^{2},
\end{equation}
the FS is finally obtained from (\ref{eq:fidelity_suspt_trY}) as
\begin{eqnarray}
\chi_f=-2\mathrm{Tr}\mathcal{Y} & = & \frac{1}{4}F,
\end{eqnarray}
where F is exactly the expression of QFI for a non-full rank density matrix~\cite{Liu2013}
\begin{eqnarray}
F & = & \sum_{i=1}^{M}\frac{\left(\partial_{\theta}\lambda_{i}\right)^{2}}{\lambda_{i}}+\sum_{i=1}^{M}4\lambda_{i}\langle\partial_{\theta}\psi_{i}|\partial_{\theta}\psi_{i}\rangle-\sum_{i,k=1}^{M}\frac{8\lambda_{i}\lambda_{k}}{\lambda_{i}+\lambda_{k}}|\langle\psi_{i}|\partial_{\theta}\psi_{k}\rangle|^{2}.
\end{eqnarray}
From this result one can find that for non-full rank density matrices, the proportional relation between QFI and fidelity susceptibility is still valid.
One should notice that the calculation
above also covers the full rank case when choosing $M=N$. Therefore, we can reach the final conclusion
that fidelity susceptibility is proportional to the quantum
Fisher information for a general density matrix.

\subsection{Application to X states}

To see how to calculate the QFI or FS, we take the X state as an example, which is defined as~\cite{Rau,Xstate,Xstate-1,Xstate-2,Xstate-3}
\begin{equation}
\rho_{X}=\left(\begin{array}{cccc}
a & 0 & 0 & w^{*}\\
0 & b & z^{*} & 0\\
0 & z & c & 0\\
w & 0 & 0 & d
\end{array}\right).\label{eq:x_state}
\end{equation}
This type of states include maximally entangled Bell states and Werner
states. The properties of this state have been widely discussed~\cite{Rau,Xstate,Xstate-1,Xstate-2,Xstate-3}.
Here we set $z=z^{*}=\sqrt{bc}$. Then the four eigenvalues of
(\ref{eq:x_state}) become
\begin{equation}
\lambda_{1}=b+c,~~ \lambda_{2}=0, ~~\lambda_{\pm}=\frac{1}{2}\left[a+d\pm\sqrt{\Delta}\right],
\end{equation}
where $\Delta=(a-d)^{2}+4|w|^{2}$. Obviously, the
dimension of the support is $M=3$. In addition, the eigenstates corresponding to $\lambda_{1}$ and $\lambda_{\pm}$ are
\begin{equation}
|\psi_{1}\rangle=\epsilon_{1}\left(0,\sqrt{\frac{b}{c}},1,0\right)^{\mathrm{T}},~~
|\psi_{\pm}\rangle=\epsilon_{\pm}\left(\frac{a-d\pm\sqrt{\Delta}}{2w},0,0,1\right)^{\mathrm{T}},
\end{equation}
where $\epsilon_{1}=\sqrt{c/(b+c)}$ and $\epsilon_{\pm}=\frac{\sqrt{2}|w|}{\sqrt{\Delta\pm(a-d)\sqrt{\Delta}}}$.

We consider an estimation of the parameter $\theta$
introduced by the following unitary operation
\begin{equation}
U=\exp\left(-i\alpha \sigma^{\alpha}_{z}\right),
\end{equation}
where $\sigma^{\alpha}_{z}=\sigma_{z}\otimes\mathbb{I}$.
Here $\mathbb{I}$ is the $2\times2$ identity matrix and $\sigma_{z}$
is a Pauli matrix, which reads $\sigma_{z}=\mathrm{diag}(1,-1)$.
In this case, the QFI reduces to
\begin{eqnarray}
F&=& 4\lambda_{\pm}\langle \Delta^{2} \sigma^{\alpha}_{z}\rangle_{\pm}+4\lambda_{1}\langle\Delta^{2} \sigma^{\alpha}_{z} \rangle_{1}  \nonumber \\
& &
-\frac{\mbox{16\ensuremath{\lambda_{+}\lambda_{-}}}}{\lambda_{+}+\lambda_{-}}|\langle\psi_{+}
|\sigma^{\alpha}_{z}|\psi_{-}\rangle|^{2}-\sum_{i=\pm}\frac{16\lambda_{i}\lambda_{1}}
{\lambda_{i}+\lambda_{1}}|\langle\psi_{i}|\sigma^{\alpha}_{z}|\psi_{1}\rangle|^{2},
\end{eqnarray}
where $\langle \Delta^{2} \sigma^{\alpha}_{z}\rangle_{i}=\langle\psi_{i}|\left(\sigma^{\alpha}_{z}\right)^{2}|\psi_{i}\rangle
-\langle\psi_{i}|\sigma^{\alpha}_{z}|\psi_{i}\rangle^{2}$.
It is obvious that QFI is only constituted by the nonzero eigenvalues and the corresponding eigenstates of the density matrix (\ref{eq:x_state}), namely, QFI is only determined by the support of (\ref{eq:x_state}).

Substituting the values of $\lambda_{\pm,1}$ and $|\psi_{\pm,1}\rangle$
into above expression, the QFI can be finally simplified as
\begin{equation}
F=16\left(\frac{|w|^{2}}{a+d}+\frac{bc}{b+c}\right). \label{eq:F_X}
\end{equation}
To guarantee the positivity of the density matrix $\rho_{\mathrm{X}}$, it requires
that all the diagonal elements of $\rho_{X}$ are positive and $ad\geq|w|^{2}$. In the mean time,
we know that $b+c\geq2\sqrt{bc}$ and $a+d\geq2\sqrt{ad}$, then one can find that
\begin{equation}
F\leq8\left(|w|+\sqrt{bc}\right).
\end{equation}
Namely, the maximum QFI is $F_{\mathrm{max}}=8(|\omega|+\sqrt{bc})$, which is satisfied under the condition $a=d=|w|$ and $b=c$. This indicates that by suitably choosing the input state, one could get the maximum QFI, which gives the minimum uncertainty of the unknown parameter $\alpha$ from Eq.~(\ref{eq:CRbound}).
One of the optimal X state in this case is the bell state $|\Phi^{+}\rangle=(|00\rangle+|11\rangle)/\sqrt{2}$. Explicitly, it is
\begin{equation}
|\Phi^{+}\rangle\langle\Phi^{+}|=\frac{1}{2}\left(\begin{array}{cccc}
1 & 0 & 0 & 1\\
0 & 0 & 0 & 0\\
0 & 0 & 0 & 0\\
1 & 0 & 0 & 1
\end{array}\right).
\end{equation}
Thus, the maximum value of the QFI is $F_{\mathrm{max}}=4$.

\section{Extention to QFIM\label{sec:QFIM}}

Quantum Fisher information matrix (QFIM) is the counterpart of QFI in multiple-parameter estimations. Since QFI for a non-full rank
density matrix $\rho$ is determined by the support of $\rho$,
then it is reasonable to speculate that QFIM could also be expressed similarly. In the following, we will calculate
the specific form of the QFIM for a density matrix with arbitrary
rank and show that it is indeed determined by the support of density matrix.

\subsection{Expression of QFIM}

We start from the definition of QFIM, whose elements read~\cite{Helstrom,Holevo}
\begin{equation}
\mathcal{F}_{\alpha\beta}=\frac{1}{2}\mathrm{Tr}\left[\rho\left\{ L_{\alpha},L_{\beta}\right\} \right],
\label{eq:QFIM_definition}
\end{equation}
where the symmetric logarithmic derivative (SLD) $L_{m}$ for the parameter
$\theta_{m}$ is determined by
\begin{equation}
\frac{\partial\rho}{\partial\theta_{m}}=\frac{1}{2}\left(\rho L_{m}+L_{m}\rho\right).\label{eq:SLD}
\end{equation}
As the same as the above section, we denote the dimension of the density matrix's support
as $M$, and the total dimension of it is $N$. And we define $[L_{m}]_{ij}:=\langle\psi_{i}|L_{m}|\psi_{j}\rangle$.
From the spectral decomposition of density matrix $\rho$ in (\ref{eq:rho_decomposition}), one can obtain the $m$th
SLD as
\begin{eqnarray}
[L_{m}]_{ij}  =
\left\{
  \begin{array}{ll}
    \frac{2\delta_{ij}\partial_{\theta_{m}}\lambda_{i}}{\lambda_{i}+\lambda_{j}}+\frac{2(\lambda_{j}
    -\lambda_{i})}{\lambda_{i}+\lambda_{j}}\langle\psi_{i}|\partial_{\theta_{m}}\psi_{j}\rangle, \label{eq:L_m} & \hbox{$i,j\in[1,M]$}; \\
    $arbitrary~value$, & \hbox{others}.
  \end{array}
\right.
\end{eqnarray}
Here $[L_{m}]_{ij}$ could be an arbitrary value out of the support of the density matrix.
However, this arbitrariness has no influence on the determinacy of QFIM. This
is because these random values are not involved in the calculation,
which will be shown below.

Based on the definition (\ref{eq:QFIM_definition}), the elements of QFIM can be expressed by
\begin{eqnarray}
\mathcal{F}_{\alpha\beta} = \frac{1}{2}\sum_{i=1}^{M}\sum_{j=1}^{N}\lambda_{i}\left([L_{\alpha}]_{ij}[L_{\beta}]_{ji}
+[L_{\beta}]_{ij}[L_{\alpha}]_{ji}\right), \label{eq:fff}
\end{eqnarray}
where the identity $\sum_{i=1}^{N}|\psi_{i}\rangle\langle\psi_{i}|=\mathbb{I}$
has been used. From Eq.~(\ref{eq:L_m}), one find that when $i\in[1,M]$
and $j\in[1,N]$,
\begin{equation}
[L_{\alpha}]_{ij}[L_{\beta}]_{ji}
=\frac{4\left(\lambda_{i}-\lambda_{j}\right)^{2}}{\left(\lambda_{i}+\lambda_{j}\right)^{2}}\langle
\partial_{\alpha}\psi_{i}|\psi_{j}\rangle\langle\psi_{j}|\partial_{\beta}\psi_{i}\rangle
+\frac{4\left(\partial_{\alpha}\lambda_{i}\right)\left(\partial_{\beta}\lambda_{j}\right)\delta_{ij}}
{\left(\lambda_{i}+\lambda_{j}\right)^{2}},
\end{equation}
with $\partial_{\alpha,\beta}$ the logogram of $\partial_{\theta_{\alpha,\beta}}$.
For a fixed $i$ satisfying $i\leq M$, there is
\begin{equation}
\sum_{j=M+1}^{N}[L_{\alpha}]_{ij}[L_{\beta}]_{ji}= 4\langle\partial_{\alpha}\psi_{i}|\partial_{\beta}\psi_{i}\rangle-
\sum_{j=1}^{M}4\langle\partial_{\alpha}\psi_{i}|\psi_{j}\rangle\langle\psi_{j}|\partial_{\beta}
\psi_{i}\rangle.
\end{equation}
Then substituting above equation into Eq.~(\ref{eq:fff}), one can obtain the final expression of the element of QFIM.

As a result, the QFIM can be splitted into the summation of two parts, i.e.,
\begin{equation}
\mathcal{F}_{\alpha\beta}=F_{\mathrm{ct}}+F_{\mathrm{qt}},\label{eq:F_alpha_beta}
\end{equation}
where
\begin{equation}
F_{\mathrm{ct}}=\sum_{i=1}^{M}\frac{(\partial_{\alpha}\lambda_{i})(\partial_{\beta}\lambda_{i})}{\lambda_{i}}\label{eq:ct}
\end{equation}
is the classical contribution, which is determined by the eigenvalues of the density matrix, and
\begin{equation}
F_{\mathrm{qt}}= \sum_{i=1}^{M}4\lambda_{i}\mathrm{Re}(\langle\partial_{\alpha}\psi_{i}|\partial_{\beta}\psi_{i}\rangle)
-\sum_{i,j=1}^{M}\frac{8\lambda_{i}\lambda_{j}}{\lambda_{i}
+\lambda_{j}}\mathrm{Re}(\langle\partial_{\alpha}\psi_{i}|\psi_{j}\rangle\langle\psi_{j}|\partial_{\beta}
\psi_{i}\rangle)
\label{eq:qt}
\end{equation}
is the quantum contribution, determined by eigenvalues and eigenstates simultaneously.
This division between the classical and quantum contribution is similar to the case of the single-parameter
estimations~\cite{Liu2013,Paris09}.

From Eqs.~(\ref{eq:ct}) and (\ref{eq:qt}), one see that there are several properties for QFIM.
First, it is a real symmetric
matrix, i.e., $\mathcal{F}_{\alpha\beta}\in\mathbb{R}$ and $\mathcal{F}_{\alpha\beta}=\mathcal{F}_{\beta\alpha}$.
Second, like the QFI in Sec.~2, the QFIM is determined by the support of the density matrix. Moreover, the diagonal term of QFIM reads
\begin{eqnarray}
\mathcal{F}_{\alpha\alpha} & = & \sum_{i=1}^{M}\frac{(\partial_{\alpha}\lambda_{i})^{2}}{\lambda_{i}}
+\sum_{i=1}^{M}4\lambda_{i}\langle\partial_{\alpha}\psi_{i}|\partial_{\alpha}\psi_{i}\rangle
-\sum_{i,j=1}^{M}\frac{8\lambda_{i}\lambda_{j}}{\lambda_{i}
+\lambda_{j}}|\langle\partial_{\alpha}\psi_{i}|\psi_{j}\rangle|^{2}, \label{eq:F_alpha_alpha}
\end{eqnarray}
which is exactly the QFI expression for the single parameter $\theta_{\alpha}$.
In addition, for a pure state $|\psi\rangle\langle\psi|$, the
expression of QFIM reduces to the well-known result~\cite{Helstrom,Holevo}
\begin{equation}
\mathcal{F}_{\alpha\beta}=4\mathrm{Re}\left(\langle\partial_{\alpha}\psi|\partial_{\beta}\psi\rangle-\langle\partial_{\alpha}\psi|\psi\rangle\langle\psi|\partial_{\beta}\psi\rangle\right).
\end{equation}

\subsection{Application to X state}

We again take the X state (\ref{eq:x_state}) as an example. Assume
that the parametrization process is described by
\begin{equation}
U_{m}=\exp\left[-i\left(\alpha\sigma^{\alpha}_{z}+\beta\sigma^{\beta}_{z}\right)\right],
\end{equation}
here $\sigma^{\alpha}_{z}=\sigma_{z}\otimes\mathbb{I}$ and $\sigma^{\beta}_{z}=\mathbb{I}\otimes\sigma_{z}$.
We set $z=z^{*}=\sqrt{bc}$. In this case, the element of QFIM are
\begin{equation}
\mathcal{F}_{\alpha\beta}=\!\sum_{i=\pm,1}\!4\lambda_{i}\mathrm{Re}
\left(\langle\psi_{i}|\sigma^{\alpha}_{z}\sigma^{\beta}_{z}|\psi_{i}\rangle\right)
-\!\sum_{i,j=\pm,1}\!\frac{8\lambda_{i}\lambda_{j}}{\lambda_{i}
+\lambda_{j}}\mathrm{Re}\left(\langle\psi_{i}|\sigma^{\alpha}_{z}|\psi_{j}\rangle\langle\psi_{j}
|\sigma^{\beta}_{z}|\psi_{i}\rangle\right).
\end{equation}
As expected, it is only determined by the nonzero eigenvalues and the corresponding eigenstates of the density matrix.

After some calculations, the explicit form of QFIM for X state can be simplified as
\begin{equation}
\mathcal{F}=16\left[\left(\frac{|w|^{2}}{a+d}+\frac{bc}{b+c}\right)\mathbb{I}
+\left(\frac{|w|^{2}}{a+d}-\frac{bc}{b+c}\right)\sigma_{x}\right],\label{eq:QFIM_X}
\end{equation}
Here $\sigma_{x}$ is a Pauli matrix. From Eq.~(\ref{eq:QFIM_X}), one can see that its diagonal element $\mathcal{F}_{\alpha\alpha}$ is exactly the expression of QFI for single-parameter estimation shown in (\ref{eq:F_X}).

\section{Conclusion\label{sec:Conclusion}}

In this paper, we study the relationship between the fidelity susceptibility and quantum Fisher information. We give a rigorous proof that the fidelity susceptibility is determined by the support of the density matrices, and it is proportional to the quantum Fisher information. Particularly, this
proof is focused on the density matrices with non-full ranks. However, the proof can be easily extended to the
full rank case. Then we apply the result to a X state.
Furthermore, we show that, similar to the quantum Fisher information, for a non-full rank density matrix,
the quantum Fisher information matrix is also determined by the support of the density matrix.
We also take the X state as an example to apply this expression.

\ack
This work was supported by the NFRPC through Grant No. 2012CB921602
and the NSFC through Grants No. 11025527 and No. 10935010. H.~N. Xiong
acknowledges the NSFC through Grant No. 11347220.

\end{document}